\documentclass{emulateapj}

\usepackage{natbib}
\newcommand{\nar}{{\it New Astronomy Reviews}}
\newcommand{\na}{{\it New Astronomy}}

\slugcomment{Submitted Feb 28, 2012; accepted Oct 2, 2012 for publication in ApJ}

\shorttitle{Kpc-scale radio structures in NLS1s}
\shortauthors{Doi et al.}

\begin{document}

\title{Kiloparsec-scale Radio Structures in Narrow-line Seyfert 1 Galaxies}

\author{Akihiro Doi\altaffilmark{1,2}, Hiroshi Nagira\altaffilmark{3}, Nozomu Kawakatu\altaffilmark{4}, Motoki Kino\altaffilmark{1,5}, Hiroshi Nagai\altaffilmark{5}, and Keiichi Asada\altaffilmark{6}}

\altaffiltext{1}{The Institute of Space and Astronautical Science, Japan Aerospace Exploration Agency, 3-1-1 Yoshinodai, Chuou-ku, Sagamihara, Kanagawa 252-5210, Japan}\email{akihiro.doi@vsop.isas.jaxa.jp}
\altaffiltext{2}{Department of Space and Astronautical Science, The Graduate University for Advanced Studies, 3-1-1 Yoshinodai, Chuou-ku, Sagamihara, Kanagawa 252-5210, Japan}
\altaffiltext{3}{Graduate school of Science and Engineering, Yamaguchi University, 1677-1 Yoshida, Yamaguchi, Yamaguchi 753-8512, Japan}
\altaffiltext{4}{Graduate School of Pure and Applied Sciences, University of Tsukuba, 1-1-1 Tennodai, Tsukuba 305-8571, Japan}
\altaffiltext{5}{National Astronomical Observatory of Japan, 2-21-1 Osawa, Mitaka, Tokyo 181-8588, Japan}
\altaffiltext{6}{Academia Sinica Institute of Astronomy and Astrophysics, P.O. Box 23-141, Taipei 10617, Taiwan}

\begin{abstract}
%
%
We report the finding of kiloparsec~(kpc)-scale radio structures in three radio-loud narrow-line Seyfert~1~(NLS1) galaxies from the Faint Images of the Radio Sky at Twenty-centimeters~(FIRST) of the Very Large Array~(VLA),    
which increases the number of known radio-loud NLS1s with kpc-scale structures to~six, including two~$\gamma$-ray emitting NLS1s~(PMN~J0948+0022 and~1H~0323+342) detected by the {\it Fermi} Gamma-ray Space Telescope.  
The detection rate of extended radio emissions in NLS1s is lower than that in broad-line active galactic nuclei~(AGNs) with a statistical significance.  
We found both core-dominated~(blazar-like) and lobe-dominated~(radio-galaxy-like) radio structures in these six NLS1s, 
which can be understood in the framework of the unified scheme of radio-loud AGNs that considers radio galaxies as non-beamed parent populations of blazars.  
Five of the six NLS1s have (i)~extended radio luminosities suggesting jet kinetic powers of $\ga10^{44}$~erg~s$^{-1}$, 
which is sufficient to make jets escape from hosts' dense environments, 
(ii)~black~holes of $\ga10^{7}$~$\mathrm{M_\sun}$, which can generate the necessary jet powers from near-Eddington mass~accretion, and  
(iii)~two-sided radio structures at kpc~scales, requiring expansion rates of $\sim0.01c$--$0.3c$ and kinematic ages of $\ga10^7$~years.   
On the other hand, most typical NLS1s would be driven by black~holes of $\la10^{7}$~$\mathrm{M_\sun}$ in a limited lifetime of $\sim10^7$~years.   
Hence the kpc-scale radio structures may originate in a small window of opportunity during the final stage of the NLS1 phase just before growing into broad-line AGNs.  
\end{abstract} 


\section{INTRODUCTION}\label{section:introduction} 

Narrow-line Seyfert~1 galaxies~(NLS1s) belong to a class of active galactic nuclei~(AGNs) identified by their optical properties of flux ratio [\ion{O}{3}]/H$\beta<3$~and Balmer lines that are only slightly broader than forbidden lines (\citealt{Osterbrock:1985}), defined as FWHM(H$\beta)<2000$~km~s$^{-1}$ \citep{Goodrich:1989}.  It has been suggested that the unusually narrow Balmer lines and other extreme properties, such as strong permitted \ion{Fe}{2} emission lines \citep{Boroson:1992}, rapid X-ray variability \citep{Pounds:1995,Leighly:1999a}, and a steep soft X-ray spectrum \citep{Wang:1996,Boller:1996,Leighly:1999}, are related to high mass accretion rates close to the Eddington limit \citep{Boroson:1992,Brandt:1998,Sulentic:2000,Mineshige:2000} on relatively low-mass black holes \citep[$\sim 10^5$--10$^{7.5}$~M$_\sun$;][]{Peterson:2000,Hayashida:2000,Grupe:2004,Zhou:2006}.  Many studies suggest that AGNs with low-mass black holes at high accretion rates tend to be radio quiet \citep[e.g.,][]{Lacy:2001,Ho:2002,Greene:2006}; radio loudness $R$ is defined as the ratio of 5~GHz radio to {\it B}-band flux densities, with a threshold of $R=10$ separating the radio-loud and radio-quiet objects \citep{Kellermann:1989}.   In fact, only $\sim7$\% of NLS1s are actuarially radio loud ($R>10$), and $\sim2.5$\% are very radio loud NLS1s ($R>100$; \citealt{Komossa:2006}, also see \citealt{Zhou:2006}).  Thus, radio jet activities on NLS1s are generally weak.  NLS1 radio sources are considered to be compact \citep[$\la 300$~parsec;][]{Ulvestad:1995} even for radio-loud objects unresolved at $\sim5\arcsec$ resolution of the Faint Images of the Radio Sky at Twenty-Centimeters \citep[FIRST;][]{Becker:1995} survey using the Very Large Array (VLA) \citep{Zhou:2006,Whalen:2006,Yuan:2008}.  Hence, the capability of large-scale radio jet activities in NLS1s has remained unknown.  

The unified scheme of radio-loud AGNs postulates that blazars are pole-on-viewed counterparts of radio galaxies \citep{Antonucci:1984,Antonucci:1985,Urry:1995}.  Several NLS1s have been recently detected by the Large Area Telescope onboard the {\it Fermi} Gamma-ray Space Telescope \citep{Abdo:2009a,Abdo:2009,Foschini:2011}; it has been argued that these NLS1s represent the third class of $\gamma$-ray emitting AGNs along with blazars and nearby radio galaxies.  The $\gamma$-ray detections strongly suggest the presence of blazar-like phenomena as a result of relativistic beaming effects due to a fast jet aligned close to our line of sight \citep{Blandford:1979}.  Moreover, the presence of highly relativistic jets has been implied in radio-loud NLS1s on the basis of radio flux variability \citep{Zhou:2003,Doi:2006,Doi:2011a,Yuan:2008}, very-long baseline interferometry~(VLBI) imaging \citep{Doi:2006,Doi:2011a,Giroletti:2011}, and modeling of spectral energy distributions \citep{Zhou:2007,Yuan:2008,Abdo:2009}.  An enormous amount of kinetic power of the relativistic jets is expected to be dissipated into thermal reservoirs and radiation at radio lobes \citep{Scheuer:1974,Begelman:1989}.  Finding extended radio structures in radio-loud and $\gamma$-ray emitting NLS1s is crucial for understanding the jet activity of the NLS1 class in the framework of the unified scheme of radio-loud AGNs.

Only three radio-loud NLS1s are known to display kiloparsec~(kpc)-scale radio morphology: 
\object[FBQS J1644+2619]{FBQS~J1644+2619} with a core plus one emission component separated by 30~kpc in the FIRST image \citep{Doi:2011a,Whalen:2006}, 
\object[1H 0323+342]{1H~0323+342} with a core plus $\sim 15$-kpc two-sided jet structures \citep{Anton:2008}, 
and \object[PKS 0558-504]{PKS~0558$-$504} with a core plus $\sim 46$-kpc two-sided radio structures \citep{Gliozzi:2010}.  
These characteristics may offer glimpse into the evolution of large-scale radio structures in the rapid-growth phase for supermassive black holes at the lower end of the AGN mass function \citep{Mathur:2000,Kawaguchi:2004}.  
incidentally, \citet{Komossa:2006} suggested that radio-loud NLS1s are actually found at the high-mass end among the NLS1 population in a sample.     
%
%
%
%

In this paper, we report the finding of kpc-scale radio structures in three additional NLS1s from VLA FIRST images.  With these detections, the number of known radio-loud NLS1s with kpc-scale radio structures increases to six.  Furthermore, we discuss the connection between the jet activity and the growth of black hole on the basis of these six NLS1 radio sources.  
In Section~\ref{section:data} of this paper, we describe the procedures of searches for extended radio emissions and the data reductions of additional VLA archival data.  In Section~\ref{section:results}, we report radio structures and the statistics of detection rate.  Section~\ref{section:discussion} comprises to the discussions on physical origins of the observed radio structures, black hole masses, and kinematic ages for the NLS1 radio sources.  Finally, we summarize our study in Section~\ref{section:summary}.  Throughout this paper, a $\Lambda$CDM cosmology with $H_0=71$~km~s$^{-1}$~Mpc$^{-1}$, $\Omega_\mathrm{M}=0.27$, and $\Omega_\mathrm{\Lambda}=0.73$ is adopted \citep{Komatsu:2009}.

\section{Data Analyses}\label{section:data}

\subsection{Extended Radio Emissions Search}\label{section:FIRSTsearch}

We searched for extended radio emissions by using the VLA FIRST images (version Jul.~16,~2008) at 1.4~GHz with a $\sim$5\arcsec~resolution around NLS1s from three previously reported samples, including those of (i)~\citet{Zhou:2006}, who presented 2011 optical-selected NLS1s at $z\la0.8$ from the Sloan Digital Sky Survey (SDSS) 3rd data release \citep[DR3;][]{Abazajian:2005}; (ii)~\citet{Yuan:2008}, who presented 23 optically-selected, very radio-loud ($R>100$) NLS1s from the updated version of \citet{Zhou:2006} by using SDSS 5th data release \citep[DR5;][]{Adelman-McCarthy:2007} and FIRST images; and (iii)~\citet{Whalen:2006}, who presented 62 radio-selected NLS1s at $z=$0.065--0.715 from the FIRST Bright Quasar Survey \citep[FBQS;][]{White:2000}.  Some fractions of objects are listed redundantly.  Because the conventional NLS1 definition of FWHM(H$\beta)<2000$~km~s$^{-1}$ was not strictly applied in any of the three NLS1 catalogs (see original papers for details), we applied the definition in this paper to obtain 1784, 17, and 47 objects from the samples of (i), (ii), and (iii), respectively.

We started searching for resolved radio structures by selecting FIRST-cataloged radio sources on the basis of two criteria: (1)~position, $<$10\arcsec\ in radius from the optical positions and (2)~intensity, $>$3~mJy~beam$^{-1}$ in a typical image noise of 0.15~mJy~beam$^{-1}$.  To the FIRST images for the selected sources, we imposed two additional criteria: (3)~the presence of extended structures associated with FIRST-cataloged radio emission determined through visual inspection and (4)~the absence of relation to any possible optical or infrared counterpart for extended emission, according to the NASA/IPAC Extragalactic Database~(NED).  As a result, we detected three sources: \object[PMN J0948+0022]{PMN~J0948+0022}, \object[SDSS J145041.93+591936.9]{SDSS~J145041.93+591936.9}, and \object[FBQS J1644+2619]{FBQS~J1644+2619}.  These radio sources appear to be unresolved central components that are accompanied by extended emissions.     

To prevent neglecting coreless or weak-core radio galaxies and large radio galaxies with isolated radio lobes at sites distant from host galaxies, we also inspected FIRST cut-out images centered at the optical positions of all NLS1 samples with the following parameters: (1)~field of view, $5\times5$~arcsec$^2$; (2)~intensity, $>$3~mJy~beam$^{-1}$; (3)~structure, emissions symmetrically located with respect to the central position or an isolated jet/lobe-like feature pointing back to the central position as determined through visual inspection; and (4)~no relation to any possible optical or infrared counterpart, according to NED.  As a result, we detected one source, \object[SDSS J120014.08-004638.7]{SDSS~J120014.08$-$004638.7}, which appears to have symmetrically located emissions without a central component.  

Here we mention cases of suspected emissions that have been subsequently exonerated.  We found isolated emissions separated by $\sim38\arcsec$\ and $\sim66\arcsec$\ from the unresolved cores of FBQS J075800.0+392029 and FBQS J1448+3559, respectively; no optical/infrared counterparts were detected.  However, we determined that these emissions are unrelated because deeper images at higher spatial resolutions by our analyses of VLA archival data at 1.4 and 5~GHz (AL450, AG151, and AB982) showed no suggestive radio structure.  
A radio source FIRST~J094901.5+002258 without an optical/infrared counterpart was located at a separation of 71\arcsec\ at a position angle of 62\degr\ from \object[PMN J0948+0022]{PMN~J0948+0022}, which is one of the detected NLS1s.  This source was previously reported out by \citet{Komossa:2006} and \citet{Foschini:2010}, who recognized it as an unrelated source; we determined the same result on the basis of our analyses of archival VLA data (AD489 and AK360).

\begin{table*}
\begin{center}
\caption{Properties of NLS1s with kpc-scale radio structures\label{table1}}
\begin{tabular}{lcccccccccc}
\hline
\hline
Galaxy name & $z$ & FWHM(H$\beta$) & [O~$_\mathrm{III}$]/H$\beta$ & $R_{4570}$ & $\Gamma_\mathrm{softX}$ & $S_\mathrm{1.4GHz}^\mathrm{FIRST}$ & $f_B$ & $\log{L_\mathrm{X}}$ & $\log{M_\mathrm{BH}}$ & Ref. \\
 &  & (km~s$^{-1}$) &  &  &  & (mJy) & (mJy) & (erg s$^{-1}$) & (M$_\mathrm{\sun}$) \\
(1) & (2) & (3) & (4) & (5) & (6) & (7) & (8) & (9) & (10) & (11) \\
\hline
\object[PMN J0948+0022]{PMN~J0948+0022} & 0.5846 & 1432 & 0.10 & 1.22 & 2.26 & 111.5 & 0.13 & 45.5 & 7.5 & 1,2,3 \\
\object[SDSS J145041.93+591936.9]{SDSS~J145041.93+591936.9} & 0.2021 & 1159 & 0.49 & 0.7 & \ldots & 3.4 & 0.08 & \ldots & 6.5 & 2,4 \\
\object[SDSS J120014.08-004638.7]{SDSS~J120014.08-004638.7} & 0.1794 & 1945 & 0.30 & 0.12 & \ldots\tablenotemark{a} & 27.1 & 0.16 & 43.9 & 7.4 & 4,5 \\
 \\
\object[FBQS J1644+2619]{FBQS~J1644+2619} & 0.1443 & 1507 & 0.11 & 0.75 & 2.19 & 108.2 & 0.16 & 43.4 & 6.9 & 1 \\
\object[1H 0323+342]{1H~0323+342} & 0.0629  & 1520 & 0.12 & 2.0 & 2.02 & 614.3\tablenotemark{b} & 3.84 & 43.9 & 7.3 & 6 \\
\object[PKS 0558-504]{PKS~0558$-$504} & 0.1372 & 1250 & 0.12 & 1.56 & 2.99 & 184\tablenotemark{c} & 3.79 & 44.6 & 7.8 & 7,8,9,10 \\
\hline
\end{tabular}
\end{center}
\tablecomments{The former three sources are first detections of extended radio structures in NLS1s; the latter objects are previously known sources.  Column~1: galaxy name; Column~2: redshift; Column~3: line width of H$\beta$ of broad component; Column~4: flux ratio of [\ion{O}{3}] to H$\beta$; Column~5: flux ratio of \ion{Fe}{2} multiplets in the range 4434--4684$\mathrm{\AA}$ to H$\beta$ \citep{Veron-Cetty:2001a}; Column~6: soft X-ray photon index; Column~7: total flux density in FIRST image, except for FBQS~J1644+2619, determined on the basis of archival data (AP0501; Figure~\ref{figure2}); Column~8: $k$-corrected $B$-band optical flux density, calculated from $B$-band magnitude listed in \citet{Veron-Cetty:2010}, Galactic extinction, and an assumed optical spectral index $\alpha=-0.5$; Column~9: 2--10~keV X-ray luminosity (the value for \object[SDSS J145041.93+591936.9]{SDSS~J145041.93+591936.9} was not found in previous studies); Column~10: black hole mass estimated from the virial relationship of a broad-line component (Section~\ref{section:escape-to-kpc}); and Column~11: references of optical and X-ray properties and black hole mass.}
\tablenotemark{a}{Absorption-corrected photon index was not available in previous studies.  We assumed $\Gamma=2.2$ to estimate $\log{L_\mathrm{X}}$.}

\tablenotemark{b}{Total flux density from NVSS.}

\tablenotemark{c}{Radio image at 4.8~GHz and its radio properties presented by \citet{Gliozzi:2010} were converted to 1.4-GHz flux densities assuming $\alpha=-0.3$ and $-0.7$ for core and lobes, respectively \citep{Gliozzi:2010}.}
\tablerefs{
(1)~\citealt{Yuan:2008},
(2)~\citealt{Zhou:2006},
(3)~\citealt{Abdo:2009b},
(4)~\citealt{Greene:2007},
(5)~\citealt{Anderson:2007},
(6)~\citealt{Zhou:2007},
(7)~\citealt{Corbin:1997},
(8)~\citealt{Remillard:1986},
(9)~\citealt{Gliozzi:2010},
(10)~\citealt{Papadakis:2010}.
}
\end{table*}

\subsection{Archival VLA Data of FBQS~J1644+2619}\label{section:J1644analyses}

For \object[FBQS J1644+2619]{FBQS~J1644+2619}, one of our detected sources, archival VLA data of deeper imaging at a spatial resolution higher than that of the FIRST image were available (Project code: AP0501), which allowed us to perform more detailed investigations.  We retrieved and reduced all the available archival VLA data\footnote{AB0611 (July~27, 1991, VLA-A at 8.4~GHz), AK0360 (May~07, 1994, VLA-AB at 4.9~GHz), AL0485 (April~13, 1999, VLA-D at 8.4~GHz), AM0593 (May~22, 1998, VLA-A at 8.4~GHz), and AP0501 (February~07, 2006, VLA-A at 1.4~GHz; January~15, 2007, VLA-C at 1.4~GHz)} obtained at relatively high spatial resolutions for FBQS~J1644+2619.  Data reduction and imaging procedures were performed by using the {\tt Astronomical Image Processing System (AIPS)} according to standard operating procedures.  The AP0501 data obtained by using the A-array configuration at 1.4~GHz provided the deepest image of rms noise of $1\sigma=90\ \mu$Jy~beam$^{-1}$ at a resolution of 1\farcs5.  Flux densities of each component were measured by Gaussian-profile fitting with the {\tt AIPS} task {\tt JMFIT}.  Errors were determined by root-sum-square of the fitting error and an amplitude calibration error of 5\%.

\section{RESULTS}\label{section:results}
We identified four NLS1 radio sources with kpc-scale structures from the FIRST image searches described in Section~\ref{section:FIRSTsearch}.  

\begin{figure*}
\epsscale{1.15}
\plotone{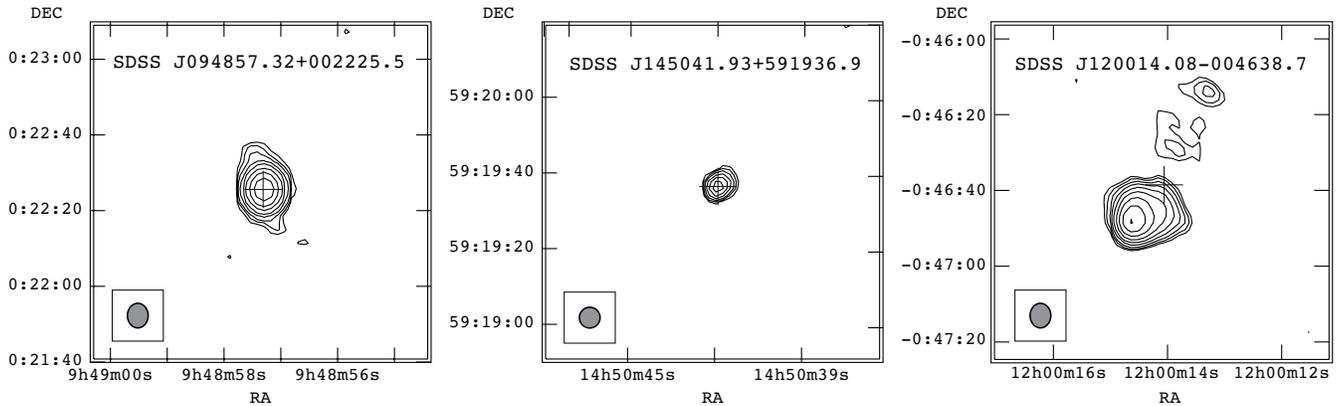}
\figcaption{VLA 1.4~GHz FIRST images of resolved NLS1s radio sources.  Contour levels are separated by factors of $\sqrt{2}$~(2 for PMN~J0948+0022) beginning at $3\sigma$ of the rms~noise ($1\sigma=0.154$, 0.137, and 0.145~mJy~beam$^{-1}$ for PMN~J0948+0022, SDSS~J145041.93+591936.9, and SDSS~J120014.08-004638.7, respectively).  Imaged region is $1\farcs5 \times 1\farcs5$.  The cross indicates the optical position of SDSS with a $10\sigma$ error ($1\sigma=0\farcs5$).\label{figure1}}
\end{figure*}

The radio structures in \object[PMN J0948+0022]{PMN~J0948+0022} and \object[SDSS J145041.93+591936.9]{SDSS~J145041.93+591936.9} are newly discovered.  The third source, \object[SDSS J120014.08-004638.7]{SDSS~J120014.08$-$004638.7}, is first identified as an NLS1 radio source in the present paper, although \citet{Zamfir:2008} previously mentioned this source in their study of SDSS quasars in a table outlining the presence of extended structures.  Thus, these three sources (Figure~\ref{figure1}) increase the number of known radio-loud NLS1s with kpc-scale radio structures to six (Table~\ref{table1}).

Although the fourth source, \object[FBQS J1644+2619]{FBQS~J1644+2619}, is a previously known NLS1 exhibiting extended radio emission in the FIRST image \citep{Doi:2011a,Whalen:2006}, its detailed structure with a deeper and higher angular resolution VLA image is first presented in the present paper [Figure~\ref{figure2} (B)].  
  
The observed radio structures are described in the following subsections, and their radio properties are listed in Table~\ref{table2}, including radio morphology, radio structure size, flux densities of the central (core) and of the extended components (lobes), and the core dominance parameter $r^\mathrm{core/lobe}_\mathrm{obs.}$, which is defined as the flux ratio of the core to extended emissions.

\subsection{Individual Objects}\label{section:individualobject}
\subsubsection{New detections of kpc-scale radio structures in NLS1s}\label{section:newdetection}
\paragraph{PMN~J0948+0022}
The FIRST image clearly shows two-sided elongations of northern (2.2~mJy) and southern (0.8~mJy) components centered at the core (108.9~mJy), i.e., a core-dominated structure.  The northern extent corresponds to 52~kpc in projected distance.  The direction of the northern extent is consistent with that of one-sided jets at pc scales in VLBI images \citep{Doi:2006,Giroletti:2011}.

\paragraph{SDSS J145041.93+591936.9}
This source shows a core with a northeast elongation that can be modeled by using two point sources of 2.8~mJy and 0.6~mJy, located $\sim0\farcs2$ and $\sim3\farcs8$, respectively, from the SDSS position.  Because the brighter component coincides with the SDSS position with an accuracy of $\sim$0\farcs5, it might be a core; the weaker component may be a one-sided extending arm.  However, its unremarkable radio structure may constitute a first step for additional follow-ups rather than conclusive evidence for classification as an NLS1 radio galaxy.  The northeast extent corresponds to 19~kpc.  The total flux density is dominated by the core.  The core dominance parameter $r^\mathrm{core/lobe}_\mathrm{obs.}$ is 4.7 (Table~\ref{table2}).

\paragraph{SDSS J120014.08-004638.7}
In their previous study of SDSS quasars, \citet{Zamfir:2008} mentioned the presence of extended structures in the FIRST image.  However, this source is a genuine NLS1 according to conventional definitions (Table~\ref{table1}), and the FIRST image is first presented in the present paper (Figure~\ref{figure1}).  This source exhibits a significantly resolved bright emission (21.8~mJy), similar to that of a radio lobe with a hotspot, separated by 10\farcs3 southeast of the SDSS position with an accuracy of $\sim$0\farcs5; the radio emission is located at substantial distance outside the isophotal radius (5\farcs71) of the host galaxy at $r$-band at a reference level of 25.0~mag~arcsec$^{-2}$.  In addition, a counter radio feature is apparent (5.3~mJy).  \citet{Zamfir:2008} described this arrangement as a ``possible core + lobe structure.''  However, a significant positional discrepancy exists between radio and optical positions.  A putative radio core at the optical position is indistinguishable from the southeast lobe.  Based on the separation between the southeast lobe and the optical position (corresponds to 43~kpc) and the fact that the total flux density is dominated by the lobes (the core dominance parameter $r^\mathrm{core/lobe}_\mathrm{obs.}$ is $<0.03$; Table~\ref{table2}), we suggest that this system either has a weak core with $\la1$~mJy or is coreless two-sided Fanaroff--Riley class-II (FR-II) radio galaxy \citep{Fanaroff:1974}.  Of the six NLS1s with kpc-scale radio structures, this object is the only lobe-dominated source.

\subsubsection{Previously known source with kpc-scale radio structures}\label{section:previouslyknownsources}

\begin{figure*}
\epsscale{1.0}
\plotone{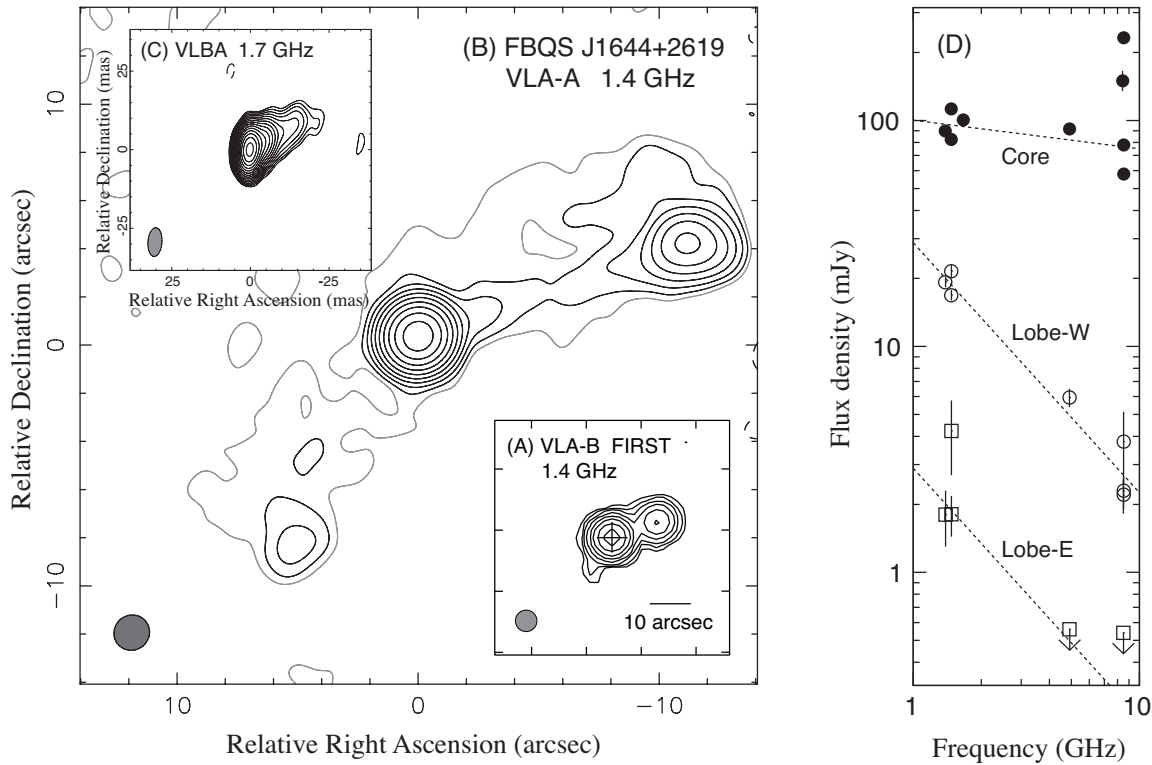}
\figcaption{Radio images and radio continuum spectra for FBQS~J1644+2619.  Contour levels are separated by factors of 2~($\sqrt{2}$ for (C)) beginning at $3\sigma$ of the rms~noise in a blank sky.  Negative and positive contours are shown as dashed and solid curves, respectively.  (A)~VLA FIRST image at 1.4~GHz.   Image rms noise is $1\sigma=0.155$~mJy~beam$^{-1}$.  Spatial resolution is 5\farcs4, as shown in the lower left-hand corner.  (B)~VLA deep image at 1.4~GHz using the A-array configuration.  An additional $2\sigma$ contour level is represented by gray curves.  Image rms~noise is $1\sigma=90\ \mu$Jy~beam$^{-1}$.  Spatial resolution is 1\farcs5 as shown in the lower left-hand corner.  (C)~VLBI image at 1.7~GHz at the core region \citep{Doi:2011a}.  (D)~Radio spectra created on the basis of all available flux-density measurements from previous studies \citep{Doi:2007,Doi:2011a} and our analyses of several archival data (Section~\ref{section:J1644analyses}).  Filled circles, open circles, and open squares represent the flux densities of the core, western lobe, and eastern lobe, respectively.  Dashed lines represent fitted power-law spectra determined by least-square method; a spectral index of $\alpha=-1.1 \pm 0.1$ was determined from the western lobes.  Because only upper limits of flux density were available at higher frequencies for the eastern lobe, the same spectral index as that for the western lobe was assumed for the fit.\label{figure2}}
\end{figure*}

\paragraph{FBQS J1644+2619}\label{section:FBQS J1644+2619}
This source appears to be a double structure of the two FIRST-cataloged components; its FIRST image was previously presented by \citet{Whalen:2006} and \citet{Doi:2011a}.  However, the deeper VLA image at a higher angular resolution clearly shows two-sided radio morphology [Figure~\ref{figure2} (B)] similar to that in typical FR-II radio galaxies, which terminates at hotspots in radio lobes.  In a simple binary classification as the ratio ($r_s$) of the separation between the brightest regions and the total size of the radio source \citep[$r_s<0.5$ for FR~I and $r_s>0.5$ for FR~II;][]{Fanaroff:1974}, $r_s \sim 0$ for FBQS~J1644+2619 should be classified as FR~I because the core dominates ($r^\mathrm{core/lobe}_\mathrm{obs.}=3.19$; Table~\ref{table2}).  However, disregarding the high core dominance (Section~\ref{section:coredominance}), both sides of the radio structure are clearly edge brightened with $r_s\sim0.9$, similar to that exhibited by FR~II rather than edge darkened as in case of FR~I.  The western component separated by 30~kpc with respect to the core is presumably a hotspot in an approaching lobe.  
The southeastern component is presumably a hotspot in a counter (receding) lobe, whose signature can be seen in the FIRST image [Figure~\ref{figure2} (A)].  

Non-simultaneous radio spectra created from all reduced archival data and literature [Figure~\ref{figure2} (D)] show steep power-law spectra of an index of $\alpha=-1.1$ ($S_\nu \propto \nu^\alpha$, where $S_\nu$ is the flux density at the frequency $\nu$) in lobes and a highly variable core.  
Variability brightness temperatures \citep[e.g.,][]{Lahteenmaki:1999} do not significantly exceed equipartition brightness temperatures \citep[$\sim10^{11}$~K;][]{Readhead:1994} or an inverse Compton limit \citep[$\sim10^{12}$~K;][]{Kellermann:1969} for the available core flux variations.

\subsection{Detection Rate of Extended Radio Sources}\label{section:fractionofNLS1RG}
In this section, we use two control samples to examine whether the detection rate of the extended radio sources in NLS1s differs from that in broad-line AGNs (hereafter refereed to as BLS1s) in a statistical sense.  The control sample of BLS1s was created on the basis of a sample presented by \citet{Rafter:2011} that found 63~sources with extended radio emissions by matching SDSS-DR4-based 8434~low-redshift type-1 AGNs \citep{Greene:2007} to FIRST images in a manner similar to our method based on visual inspection (Section~\ref{section:FIRSTsearch}).  Its criteria were (a$_\mathrm{B}$)~$z<0.35$, (b$_\mathrm{B}$)~$>1$~mJy at radio, (c$_\mathrm{B}$)~radio extended $>4\arcsec$ from optical position.  The NLS1 control sample was created on the basis of our results that used the SDSS-3DR-based 1784~NLS1s from \citet{Zhou:2006} [sample~(i); Section~\ref{section:data}].  Its criteria were (a$_\mathrm{N}$)~$z\la0.8$, (b$_\mathrm{N}$)~FWHM(H$\beta)<2000$~km~s$^{-1}$, (b$_\mathrm{N}$)~$>3$~mJy~beam$^{-1}$ at radio.  

From these two parent samples, we obtained two subsamples to meet the following common criteria: (a)~$z<0.35$, (b)~$>3$~mJy~beam$^{-1}$ at radio, (c)~radio extended $>4\arcsec$ from optical position, (d)~FWHM(H$\alpha)>1835.4$~km~s$^{-1}$ to the former parent sample for selecting BLS1s and FWHM(H$\beta)<2000$~km~s$^{-1}$ to the latter parent sample for selecting NLS1s.  In criterion~(d), the threshold of FWHM(H$\alpha)=1835.4$~km~s$^{-1}$ is equivalent to FWHM(H$\beta)=2000$~km~s$^{-1}$, according to an empirical relationship of FWHM(H$\beta)=1.07\times1000\ (\mathrm{FWHM(H\alpha})/1000\ \mathrm{km\ s}^{-1})^{1.03}$~km~s$^{-1}$ based on type-1 AGNs in SDSS DR3 at $z<0.35$ \citep{Greene:2005}.  Because the optical catalog used in \citet{Rafter:2011} includes only the values of FWHM(H$\alpha$) \citep{Greene:2007}, we used this equivalent threshold on line widths for the BLS1 control sample.      

When these criteria are strictly applied, the numbers of extended radio sources in the two control samples are 58 in 6868~BLS1s and 1 in 1000~NLS1s.  The one NLS1 with an extended radio structure is \object[SDSS J120014.08-004638.7]{SDSS~J120014.08$-$004638.7}.  \object[PMN J0948+0022]{PMN~J0948+0022} is definitely excluded by~(a).  \object[SDSS J145041.93+591936.9]{SDSS~J145041.93+591936.9} is excluded although marginally by~(c) because the radio extension is only $\sim3\farcs8$ from the optical position (Section~\ref{section:newdetection}).  \object[FBQS J1644+2619]{FBQS~J1644+2619} is excluded because it is absent in the SDSS~DR3-based NLS1 sample of \citet{Zhou:2006} [sample~(i)] and has appeared in SDSS~DR4.  Therefore, this object is listed in the SDSS~DR5-based very radio-loud NLS1 sample of \citet{Yuan:2008} [sample~(ii)].  The previously reported NLS1s with kpc-scale radio structures \object[1H 0323+342]{1H~0323+342} \citep{Anton:2008} and \object[PKS 0558-504]{PKS~0558$-$504} \citep{Gliozzi:2010} are outside the FIRST and SDSS sky coverages, respectively.  Thus, considering the two marginal cases (\object[SDSS J145041.93+591936.9]{SDSS~J145041.93+591936.9} and \object[FBQS J1644+2619]{FBQS~J1644+2619}), the number of extended radio sources could be 1--3 in 1000~NLS1s.

As a result, the detection rates of extended radio emissions are 58/6868~(0.84\%) and 1--3/1000~(0.10\%--0.30\%) in BLS1s and NLS1s, respectively.  Fisher's exact test rejects a null hypothesis that the detection rate in NLS1s is not lower than that in BLS1s with a significance level of 0.05 in any of the cases, because of probabilities of $p=0.0030$--0.0395.  Therefore, {\it the detection rate of extended radio emissions in NLS1s is lower than that in BLS1s in a statistical sense on the basis of our two SDSS--FIRST control samples at $z<0.35$}.

\section{DISCUSSION}\label{section:discussion}
With our new findings, including the tentative detection of the extended radio emission in \object[SDSS J145041.93+591936.9]{SDSS~J145041.93+591936.9}, the number of known NLS1s with kpc-scale radio structures increases to six.  The characteristics of these six objects are discussed in the following subsections.  

\subsection{Different Jet/Lobe Speeds at Different Geometric Scales}\label{section:advancingspeed}
Five of the six NLS1s exhibit two-sided structures at kpc scales [Column~(2) in Table~\ref{table2}].  On the other hand, from VLBI observations, one-sidedness in pc scales is evident in four of the five NLS1s: \object[PMN J0948+0022]{PMN~J0948+0022} \citep{Doi:2006,Giroletti:2011}, \object[1H 0323+342]{1H~0323+342} (\citealt{Lister:2005} in the MOJAVE\footnote{https://www.physics.purdue.edu/astro/mojave/} project), \object[PKS 0558-504]{PKS~0558$-$504} \citep{Gliozzi:2010}, and \object[FBQS J1644+2619]{FBQS~J1644+2619} [Figure~\ref{figure2} (C)].  The one-sided structure of \object[FBQS J1644+2619]{FBQS~J1644+2619}, which can be interpreted as a consequence of the Doppler beaming effect if jets are not intrinsically asymmetric, provided mild constraints of $\beta_\mathrm{pc}>0.74$ and $\Phi<42$\degr, where $\beta$ is the jet speed in the unit of the speed of light and $\Phi$ is the viewing angle \citep{Doi:2011a}.  In addition, a flat-spectrum and highly variable core [Figure~\ref{figure2} (D)] as well as a significant polarized radio flux ($3.45 \pm 0.46$~mJy) reported in the National Radio Astronomy Observatory VLA Sky Survey \citep[NVSS;][]{Condon:1998} indicate the presence of a blazar-like core for \object[FBQS J1644+2619]{FBQS~J1644+2619}.   The combination of one-sided morphology in pc scales and a two-sided structure in kpc scales is suggested by the concepts of beamed core and radio lobes as decelerated components.  Different viewing angles at different geometric scales without deceleration is an alternative explanation; however, it is more possible that the viewing angle remains nearly constant because the position angles of radio morphology from pc to kpc scales is constant.  Assuming intrinsically symmetric two-sided ejecta, steady jet activity, and beaming considerations \citep{Ghisellini:1993}, we can estimate the speed of a lobe of $\beta_\mathrm{kpc}=0.27$--0.36 at $\Phi<42\degr$ by using the apparent flux ratio of approaching and receding hotspots given as $R_\mathrm{F}=[(1+\beta \cos{\Phi})/(1-\beta \cos{\Phi})]^{3-\alpha}$ and by applying observed flux densities of 16.9~mJy and 1.8~mJy and a spectral index of $\alpha=-1.1$ [Figure~\ref{figure2} (D)].  It should be noted that this method assumes no variation in the luminosities of the two components during a differential time delay.  As an additional approach, on the basis of a geometric consideration of the differential time delay, we estimated $\beta_\mathrm{kpc}=0.09$--0.12 at $\Phi<42\degr$ by using the apparent extent ratio of approaching to receding lobes given as $R_\mathrm{D}=(1+\beta \cos{\Phi})/(1-\beta \cos{\Phi})$ and by applying observed separations of 11\farcs8 and 10\farcs0.  Therefore, a deceleration in $\beta$ is required from the fast ($\beta_\mathrm{pc}>0.74$) jets in pc scales to the significantly slower ($\beta_\mathrm{kpc} = 0.09$--$0.36$) radio lobes in kpc scales; a stronger constraint on $\Phi$ is obtained in Section~\ref{section:coredominance}.

\begin{table*}
\begin{center}
\caption{Radio Properties and Core Dominance Parameters\label{table2}}
{\scriptsize  
\begin{tabular}{lcccccccccccc}
\hline
\hline
Galaxy name & Morph. & $L^\mathrm{proj.}$ &  &  & $S^\mathrm{core}_\nu$ & $S^\mathrm{lobe}_\nu$ & $r^\mathrm{core/lobe}_\mathrm{obs.}$ & $r^\mathrm{core/lobe}_\mathrm{int.G01}$ & $r^\mathrm{core/lobe}_\mathrm{int.M03}$ & $\log{R_\mathrm{obs.}}$ & $\log{R_\mathrm{int.}}$ & $\log{L_\mathrm{j}}$ \\
 &  & (arcsec) & (kpc) & ($10^9$Rs) & (mJy) & (mJy) & (erg s$^{-1}$)\\
(1) & (2) & (3) & (4) & (5) & (6) & (7) & (8) & (9) & (10) & (11) & (12) & (13) \\
\hline
\object[PMN J0948+0022]{PMN~J0948+0022} & 2S & 7.9 & 52 & 17 & 108.4  & 3.0  & 36.1  & 0.023  & 0.12  & 2.73 & 1.08 & 44.3 \\
\object[SDSS J145041.93+591936.9]{SDSS~J145041.93+591936.9} & 1S? & 4.0 & 19 & 59 & 2.8 & 0.6 & 4.7  & 0.22  & \ldots & 1.33 & 0.62 &42.8 \\
\object[SDSS J120014.08-004638.7]{SDSS~J120014.08-004638.7} & 2S & 10.3 & 43 & 16 & $<1.0$ & 32.4  & $<0.03$ & 0.051  & 0.049  & 1.64 & 1.95 & 44.0 \\
 \\
\object[FBQS J1644+2619]{FBQS~J1644+2619} & 2S & 11.8 & 30 & 38 & 82.4 & 25.8 & 3.19  & 0.046  & 0.006  & 2.39 & 1.63 & 44.0 \\
\object[1H 0323+342]{1H~0323+342}\tablenotemark{a} & 2S & 20.1 & 24 & 8 & 421  & 198  & 2.1  & 0.03  & 0.02  & 2.39 & 1.38 & 44.1 \\
\object[PKS 0558-504]{PKS~0558$-$504}\tablenotemark{b} & 2S & 7.0 & 17 & 3 & 130  & 53  & 2.4 & 0.06  & 0.17  & 1.42 & 0.85 & 44.2 \\
\hline
\end{tabular}
}
\end{center}
\tablecomments{Column~1: galaxy name; Column~2: radio morphology (2S and 1S represent two-sided and one-sided structures, respectively); Columns~3--5: projected extent on one side of radio structure, in arcsec, kpc, and Schwarzschild radius, respectively; Columns~6 and 7: observed core and lobe flux densities in the FIRST image, respectively, based on archival data, with the exception of FBQS~J1644+2619 (AP0501; Figure~\ref{figure2}); Column~8: observed core--lobe flux ratio; Column~9: core--lobe flux ratio expected from empirical correlation between cores and total radio powers in radio galaxies \citep{Giovannini:2001}, by assuming spectral indices of $\alpha=0$ and $-0.7$ for core and lobes, respectively, if observed values are not available (Section~\ref{section:coredominance}); Column~10: core--lobe flux ratio expected from empirical correlation among core radio luminosity, X-ray luminosity, and black hole mass \citep{Merloni:2003} (Section~\ref{section:coredominance}); Column~11: observed radio loudness, defined as the ratio of 5~GHz radio to {\it B}-band flux densities with a threshold of $R=10$ separating radio-loud and radio-quiet objects \citep{Stocke:1992}; and Column~12: intrinsic radio loudness using the observed lobe and the intrinsic core luminosities on the basis of $r^\mathrm{core/lobe}_\mathrm{int.M03}$; that for \object[SDSS J145041.93+591936.9]{SDSS~J145041.93+591936.9} was determined on the basis of $r^\mathrm{core/lobe}_\mathrm{int.G01}$ instead; Column~13: jet kinetic power, derived from the radio luminosity of lobe component using an empirical relation between the 1.4-GHz radio luminosity and the cavity power in X-ray emitting hot gas \citep{Cavagnolo:2010}.}  

\tablenotemark{a}{Radio image at 1.4~GHz presented by \citet{Anton:2008}.  The radio properties were determined on the basis of our analyses of the same data from the VLA archive.  Radio spectral indices are $\alpha=+0.1$ and $-0.85$ for core and lobes, respectively; the core spectrum were derived from our analyses of VLBA archival data (project code: BE042) obtained by simultaneous observations at 2.3 and 8.4 GHz.} 

\tablenotemark{b}{Radio image at 4.8~GHz and its radio properties presented by \citet{Gliozzi:2010}, which were converted to 1.4-GHz flux densities assuming $\alpha=-0.3$ and $-0.7$ for core and lobes, respectively \citep{Gliozzi:2010}.}
\end{table*}

Similarly, for \object[PMN J0948+0022]{PMN~J0948+0022}, $\beta_\mathrm{pc}>0.76$ and $\Phi<22\degr$ were determined from the very high brightness temperature of core-dominated one-sided jets in the VLBI image.  On the contrary, the apparent flux ratio $R_\mathrm{F}$ of the northern (2.2~mJy) and southern (0.8~mJy) components in kpc scales (Section~\ref{section:newdetection}) requires a deceleration down to $\beta_\mathrm{kpc}=0.13$--0.15 assuming $\alpha=-0.7$ and the same range of $\Phi$ at the pc scales; a stronger constraint on $\Phi$ is obtained in Section~\ref{section:coredominance}.  Significant radio flux variability, polarized radio emissions, flat/inverted radio spectra \citep{Abdo:2009b}, and $\gamma$-ray detections \citep{Abdo:2009a} have also been reported for \object[PMN J0948+0022]{PMN~J0948+0022}.  \object[1H 0323+342]{1H~0323+342} is also a $\gamma$-ray-emitting NLS1 \citep{Abdo:2009} in the NLS1s with kpc-scale radio structures.  These observed properties indicate the presence of blazar-like beamed jets at pc scales in these NLS1s, whereas kpc-scale ones are rarely beamed.

\subsection{Core Dominances in the Unified Scheme}\label{section:coredominance}

Except the lobe-dominated case of \object[SDSS J120014.08-004638.7]{SDSS~J120014.08$-$004638.7}, all NLS1s with kpc-scale radio structures exhibit a core with significantly higher luminosity than that of extended emissions [Column~(8) in Table~\ref{table2}].  These core dominances are unusually high compared to typical radio galaxies, which are lobe-dominated with respect to cores \citep[the median values are 0.022 for FR-I and 0.003 for FR-II radio galaxies;][]{Morganti:1997}.  Similarly, blazars frequently show a prominent core that accompanies the extended radio structures of low brightness in deep VLA images \citep{Murphy:1993,Cassaro:1999,Landt:2006,Kharb:2010}.  
The possible physical origin of the high core dominances is Doppler boosting only in cores (Section~\ref{section:advancingspeed}), unless the jet activity has undergone extreme variations.  In this subsection, we determine the Doppler factors required from the observed core dominances of these NLS1s.  In addition, we constrain jet speeds and viewing angles at pc scales (Sections~\ref{section:coredominance}) 
and estimate deprojected sizes and kinematic ages of kpc-scale structures (Sections~\ref{section:escape-to-kpc} and \ref{section:ImplicationsforBHevolution}).

Two methods can be used to derive an intrinsic core power.  The first uses an empirical correlation in radio galaxies given by $\log{P_\mathrm{c}}=0.62\log{P_\mathrm{t}}+7.6$, where $P_\mathrm{c}$ is the core radio power at 5~GHz and $P_\mathrm{t}$ is the total radio power at 408~MHz in W~Hz$^{-1}$, with a scatter of 1.1~dex in $P_\mathrm{c}$ \citep{Giovannini:2001}.  Because the total radio power is measured at a low frequency and is therefore not affected by Doppler boosting, an expected intrinsic core power can be determined from the total radio power.  The second method uses the fundamental plane of black hole activity given by $\log{L_\mathrm{R}}=0.60\log{L_\mathrm{X}}+0.78\log{M_\mathrm{BH}}+7.33$, where $L_\mathrm{R}$, $L_\mathrm{X}$, and $M_\mathrm{BH}$ are the 5-GHz unbeamed core luminosity, 2--10-keV X-ray luminosity in erg~s$^{-1}$, and black hole mass in $\mathrm{M_\sun}$, respectively, with a scatter of 0.88~dex in ${L_\mathrm{R}}$ \citep{Merloni:2003}.  This relationship was determined by using the samples of X-ray binaries, nearby low-luminosity AGNs, Seyfert galaxies (including seven NLS1s), and quasars.  The key is that intrinsic core contributions expected from these two individual methods ($r^\mathrm{core/lobe}_\mathrm{int.} \sim 0.01$--0.2) are much smaller than the observed values ($r^\mathrm{core/lobe}_\mathrm{obs.} =2.1$--36.1), with the exception of \object[SDSS J120014.08-004638.7]{SDSS~J120014.08$-$004638.7} [Columns~(8)--(10) in Table~\ref{table2}].  These crucial differences can be interpreted as a Doppler beaming effect only in cores.  The some apparent discrepancies in $r^\mathrm{core/lobe}_\mathrm{int.}$ between the two methods could be due to such factors as scatters on the empirical relations, luminosity variabilities, or the uncertainties of black hole masses.

We estimated a core Doppler factor of $\delta^\mathrm{core}$ by comparing intrinsic and observed core luminosities, assuming that the core flux density is boosted by $(\delta^\mathrm{core})^{3-\alpha}$, where $\delta = \sqrt{1-\beta^2} (1-\beta\cos{\Phi})^{-1}$ \citep{Rybicki:1979}, and a flat ($\alpha=0$) spectrum if $\alpha$ is not available\footnote{$\alpha=+0.77$ for PMN~J0948+0022 \citep{Doi:2006}, $\alpha=+0.38$ for FBQS~J1644+2619 \citep{Doi:2011a}, $\alpha=+0.1$ and $\alpha=-0.3$ for 1H~0323+342 and PKS~0558$-$504, respectively (the captions in Table~\ref{table2}).}.  Subsequently, we can obtain the constraints of jet speed $\beta^\mathrm{core}$ and viewing angle $\Phi^\mathrm{core}$ for each source (Table~\ref{table3}).  Thus, all cases showing unusual core dominances require nearly pole-on-viewed jets with at least mildly or highly relativistic speeds.  It is noteworthy that these jet parameters are consistent with those obtained through SED modeling for the two $\gamma$-ray emitting NLS1s: \object[PMN J0948+0022]{PMN~J0948+0022} and \object[1H 0323+342]{1H~0323+342} \citep{Abdo:2009a,Abdo:2009}.  Therefore, the existence of radio lobes in these NLS1s is essential as places for dissipation of the huge amount of kinetic powers of the strong jets into thermal reservoirs and radiation \citep{Scheuer:1974,Begelman:1989}.  Conversely, the lobe-dominated structure of \object[SDSS J120014.08-004638.7]{SDSS~J120014.08$-$004638.7} indicates the existence of a radio galaxy in the NLS1 population as well.  {\it Thus, these NLS1s with kpc-scale radio structures can be understood in the framework of the unified scheme of radio-loud AGNs that considers radio galaxies as non-beamed parent populations of blazars} \citep{Antonucci:1984,Antonucci:1985,Urry:1995} although only a small number of objects have been studied in detail thus far.

While the detected NLS1s are very radio loud ($\log{R_\mathrm{obs.}}\ga1.5$), the intrinsic radio loudness $\log{R_\mathrm{int.}}$ based on the extended radio luminosity and the unbeamed core luminosity allows us to infer the property of their parent populations as inclined NLS1 radio sources (Table~\ref{table2}).  \object[PMN J0948+0022]{PMN~J0948+0022} and \object[PKS 0558-504]{PKS~0558$-$504} should no longer be very radio loud at the rest frame ($\log{R_\mathrm{int}} \sim 1$).  In contrast, \object[FBQS J1644+2619]{FBQS~J1644+2619}, \object[SDSS J120014.08-004638.7]{SDSS~J120014.08-004638.7}, and \object[1H 0323+342]{1H~0323+342} remain significantly radio loud at the rest frame ($\log{R_\mathrm{int.}} \ga 1.5$) because of their luminous radio lobes.  Thus, NLS1s with kpc-scale radio structures are probably provided from the populations of both intrinsically radio-intermediate and radio-loud objects rather than from only a radio-quiet population.

\subsection{Escape to kpc Scales on Low-mass Black Hole Systems}\label{section:escape-to-kpc}
The projected radio sizes on one side of the NLS1s with extended emissions are $\sim10$--50~kpc [Column~(4) in Table~\ref{table2}], which are at the smaller end of the size distribution of known radio galaxies and blazars \citep[$\sim10$--$10^3$~kpc;][and references therein]{Landt:2006} but are very large in the unit of Schwarzschild radius [Rs; Column~(5) in Table~\ref{table2}].  These sizes are equivalent to the largest known radio galaxies of several Mpc \citep[e.g., ][]{Machalski:2008}, which correspond to the several $\times 10^{10}$~Rs for assumed black hole masses of $\sim 10^{9}$~M$_\mathrm{\sun}$ typically found in radio galaxies.  Deprojected sizes are possibly larger than 100~kpc for high core dominances, i.e., nearly pole-on viewed cases [Column~(5) in Table~\ref{table3}].  Thus, their radio structures have developed into intergalactic environments outside the host galaxies.  
In this subsection, we discuss the successful escape of radio structures of the detected NLS1s to kpc scales and the cause of the apparently compact structures of most typical NLS1s.    

The majority of the NLS1s with kpc-scale structures, found so far, show FR~II-like radio lobes (\object[FBQS J1644+2619]{FBQS~J1644+2619}, \object[SDSS J120014.08-004638.7]{SDSS~J120014.08$-$004638.7}, \object[1H 0323+342]{1H~0323+342}, and \object[PKS 0558-504]{PKS~0558$-$504}) and appear to include edge-brightened lobes with $r_\mathrm{s}>0.8$ if their high core dominances are neglected.  The edge-darkened (FR~I)/edge-brightened (FR~II) morphology reflects the subsonic/supersonic speed of their lobes.  Because of strong deceleration in the nuclear dense region, an initial jet speed much faster than $0.1c$ is expected for progression up to kpc scales without disrupting hotspots \citep{Kawakatu:2008}.  Considering deceleration due to the growth of the effective cross-sectional area of lobes, the FR~I/FR~II dichotomy is determined by the jet kinetic power to ambient density ratio $L_\mathrm{j}/\bar{n}_\mathrm{a}$ \citep{Kaiser:2007,Kawakatu:2009}.  A threshold exceeding $L_\mathrm{j} /\bar{n}_\mathrm{a}=10^{44}$--$10^{45}$~erg~s$^{-1}$~cm$^{-3}$ is required to extend supersonic lobes beyond a core radius, where $L_\mathrm{j}$ is the jet kinetic power and $\bar{n}_\mathrm{a}$ is the ambient number density at the core radius ($\sim1$~kpc) \citep{Kawakatu:2009}.  
As a basic topic of discussion, considering the case of $\bar{n}_\mathrm{a} \approx 0.1$--1~cm$^{-3}$ (cf.~$n\sim0.1$~cm$^{-3}$ at the center of ellipticals; e.g., \citealt{Mathews:2003}, $n\approx1$~cm$^{-3}$ for local interstellar medium in our Galaxy; \citealt{Cox:1987}), $L_\mathrm{j}\ga10^{44}$~erg~s$^{-1}$ is required.  
We estimate $L_\mathrm{j}$ for the NLS1s with kpc-scale radio structures using their 1.4-GHz lobe luminosities according to the relationship between jet kinetic and radio powers (\citealt{Cavagnolo:2010}, see also e.g., \citealt{Willott:1999,OSullivan:2011,Merloni:2007}).  As a result, $L_\mathrm{j}\ga10^{44}$~erg~s$^{-1}$ is estimated for all sources except \object[SDSS J145041.93+591936.9]{SDSS~J145041.93+591936.9}, which shows a marginal radio structure [Column~(13) in Table~\ref{table2}].  
Thus, the observed extended radio emissions actually suggest sufficient jet kinetic powers for escaping to kpc scales in the form of supersonic lobes.

\begin{figure}
\epsscale{1.15}
\plotone{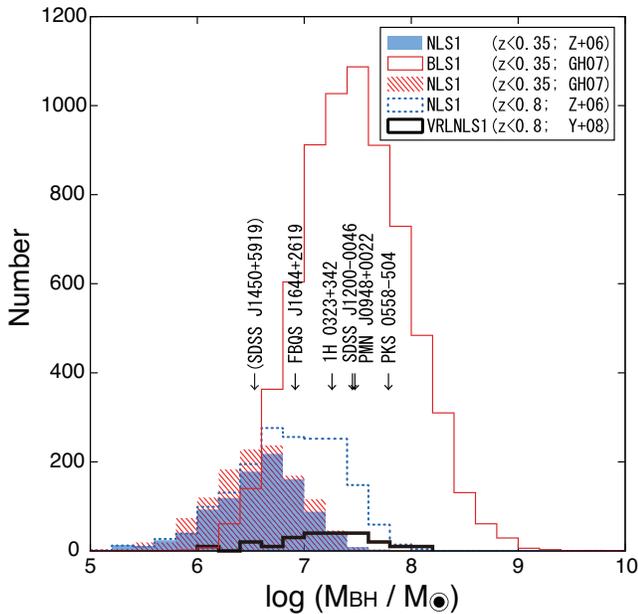}
\figcaption{{\footnotesize Histogram of black hole masses.  (1)~Shaded region: the NLS1 control sample consisting of 1000~NLS1s at $z<0.35$ from the SDSS~DR3 NLS1 sample.  (2)~Open region with thin solid line: the BLS1 control sample consisting of 6868~BLS1s at $z<0.35$ from the SDSS~DR4 type-1 AGN sample.  (3)~Hatched region: 1566~NLS1s at $z<0.35$ from the SDSS~DR4 type-1 AGN sample [see Section~\ref{section:fractionofNLS1RG} for (1)--(3)].  The two NLS1 samples display $M_\mathrm{BH} \sim 10^{5.7}$--$10^{7.4}$~$\mathrm{M_\sun}$ ($\langle \log{M_\mathrm{BH}/\mathrm{M_\sun}} \rangle = 6.6 \pm 2\sigma$).  (4)~Open region with dotted line: the sample~(i) consisting of 1784~NLS1s at $z\la0.8$ from the SDSS~DR3 NLS1 sample.  (5)~Open region with bold line: the sample~(ii) consisting of 23~very radio-loud NLS1s~(VRLNLS1) at $z\la0.8$ in the SDSS~DR5 NLS1 sample [see Section~\ref{section:FIRSTsearch} for (4)--(5)].  The number of VRLNLS1 sources was multiplied by 10 for display in this figure to compensate for the small size of the sample.  A bias to higher masses at higher $z$ due to a flux-limited characteristic of the SDSS-based sample is evident in (4) and (5).  The black hole masses of six NLS1s with kpc-scale radio structures are indicated by the positions of downward arrows.  Only PMN~J0948+0022 is at $z>0.35$.\label{figure3}}}
\end{figure}

The jet kinetic powers could be limited by the low-mass black holes of most typical NLS1s if the kinetic powers are supplied with accretion that is limited by the Eddington luminosity.  As a basic point of discussion, we offer the case of $L_\mathrm{j} \sim 0.1 L_\mathrm{Edd}$ \citep{Ito:2008}, where $L_\mathrm{Edd}=1.3\times10^{38} (M_\mathrm{BH}/M_\sun)$~erg~s$^{-1}$ is the Eddington luminosity.   
To maintain supersonic radio lobes at kpc scales, black hole masses must be $M_\mathrm{BH} \ga 10^7\ \mathrm{M_\sun}$ in order to generate $L_\mathrm{Edd} \ga 10^{45}$~erg~s$^{-1}$.  
Figure~\ref{figure3} shows histograms of black hole masses for the control samples of NLS1s, BLS1s, and other reference samples that were defined in Sections~\ref{section:FIRSTsearch} and \ref{section:fractionofNLS1RG}.  The masses have been computed by using the virial relationship of the broad-line component \citep[e.g.,][]{Kaspi:2000}; we employed a standard single-epoch procedure that uses the relationship between FWHM(H$\beta$) and optical continuum luminosity ($L_{5100\AA}$) or between FWHM(H$\alpha$) and broad-line luminosity ($L_\mathrm{H\alpha}$) given by equations~(5) or (6) in \citet{Greene:2005}.   
In the distributions,
{\it all of the NLS1s with kpc-scale radio structures are driven by relatively high-mass black holes of} $M_\mathrm{BH} \ga 10^{7}\ \mathrm{M_\sun}$, except for \object[SDSS J145041.93+591936.9]{SDSS~J145041.93+591936.9} that shows a marginal radio structure.  
These objects appear to constitute a peculiar group at the high-mass end of the distributions among the NLS1 populations and can exceed the threshold of $L_\mathrm{j}/\bar{n}_\mathrm{a}$.  
On the contrary, the black hole masses of most typical NLS1s ($\la 10^7\ \mathrm{M_\sun}$) suggest below the threshold.    
In these systems, even if jets are emanated, hotspots would be disrupted into edge-darkened (FR-I) radio lobes, which may tend to go undetected in the limited image sensitivity of FIRST \citep{Punsly:2011}, as is the case with the discovery of FR-I morphology in the radio-quiet quasar E1821+643 exclusively in a deep VLA image \citep{Blundell:2001}. 
The lower detection rate is not due to an observational limitation specific to NLS1s because the control sample of BLS1s was also investigated using FIRST images (Section~\ref{section:fractionofNLS1RG}) and shares the same difficulty in detection of FR-I morphology.
{\it Therefore, lower maximal jet kinetic powers due to lower-mass black holes may be related to the lower detection rate of the extended radio structures in the NLS1 population compared with those in the broad-line AGNs}.  
The radio emissions in most typical (radio-quiet) NLS1s are compact \citep{Ulvestad:1995} and are dominated by diffuse components confined within central regions in VLBI images (\citealt{Middelberg:2004,Orienti:2010,Giroletti:2009}; Doi~et~al.~in~prep.), which may be relevant to the link between apparent radio size and black hole mass.  
Furthermore, the link between the radio size and Eddington ratio may also be a contributor, since only a small fraction of accretion luminosity is expected to be transferred to jet power in high Eddington ratio systems such as NLS1s, according to the nonlinearity in an empirical relation that is observed in radio galaxies: $\log{L_\mathrm{j}/L_\mathrm{Edd}} = 0.49 \log{L_\mathrm{bol}/L_\mathrm{Edd}}-0.78$ \citep{Merloni:2007}.  
This may be another reason for the lower detection rate of the extended radio structures in NLS1s.

\begin{table*}
\begin{center}
\caption{Doppler-beaming Parameters Estimated from Core Dominance.  Deprojected Arm Lengths and Kinematic Ages\label{table3}}
\begin{tabular}{lccccccc}
\hline\hline
Galaxy name & $\delta^\mathrm{core}$ & $\beta^\mathrm{core}$ & $\Phi^\mathrm{core}$ & $L^\mathrm{deproj.}$ & $R_\mathrm{F}$ & $\beta^\mathrm{kpc}$ & $\log{t^\mathrm{kpc}}$ \\
 &  &  & (\degr) & (kpc) & (mJy/mJy) &  & (year) \\
(1) & (2) & (3) & (4) & (5) & (6) & (7) & (8) \\
\hline
\object[PMN J0948+0022]{PMN~J0948+0022} & 12.8 & $>0.988$ & $<4$ & $>660$ & 2.2/0.8 & $\sim0.14$ & $>7.2$ \\
\object[SDSS J145041.93+591936.9]{SDSS~J145041.93+591936.9} & 2.8 & $>0.77$ & $<21$ & $>53$ & 0.6/$<0.45$ & $>0.039$\tablenotemark{c} & $>5.7$ \\
\object[SDSS J120014.08-004638.7]{SDSS~J120014.08$-$004638.7} & \ldots\tablenotemark{a} & \ldots\tablenotemark{a} & \ldots\tablenotemark{a} & $>44$\tablenotemark{b} & 21.8/5.3 & $>0.20$\tablenotemark{b} & $>5.8$ \\
 \\
\object[FBQS J1644+2619]{FBQS~J1644+2619} & 10.9 & $>0.983$ & $<5$ & $>320$ & 16.9/1.8 & $\sim0.27$ & $>6.6$ \\
\object[1H 0323+342]{1H~0323+342} & 4.2 & $>0.892$ & $<14$ & $>100$ & 29.3/26.7 & $\sim0.012$ & $>7.4$ \\
\object[PKS 0558-504]{PKS~0558$-$504} & 2.2 & $>0.67$ & $<26$ & $>38$ & $\sim1.2$\tablenotemark{d} & $\sim0.023$\tablenotemark{d} & $>6.7$\tablenotemark{d} \\
\hline
\end{tabular}
\end{center}
\tablecomments{Column~1: galaxy name; Column~2: Doppler factor derived from observed core and intrinsic core luminosities on the basis of $r^\mathrm{core/lobe}_\mathrm{int.M03}$; that for SDSS~J145041.93+591936.9 was determined on the basis of $r^\mathrm{core/lobe}_\mathrm{int.G01}$ (Table~\ref{table2}).  Column~3: jet speed constrained by $\delta^\mathrm{core}$; Column~4: viewing angle constrained by $\delta^\mathrm{core}$; Column~5: deprojected kpc-scale arm length at the range of $\Phi^\mathrm{core}$; Column~6: apparent flux ratio of approaching to receding lobes; Column~7: speed of advance derived from the apparent flux ratio $R_\mathrm{F}$ (Section~\ref{section:advancingspeed}) at kpc scales in the range of $\Phi^\mathrm{core}$; and Column~8: kinematic age derived from $L^\mathrm{deproj.}$ assuming $\beta^\mathrm{kpc}$ as an expansion speed.}   

\tablenotemark{a}{Not available due to coreless radio structure.}  

\tablenotemark{b}{Based on a viewing angle $< 78$\degr~derived from apparent flux ratio $R_\mathrm{F}$ at kpc radio lobes.} 

\tablenotemark{c}{Only a lower limit is available due to the one-sided structure.}  

\tablenotemark{d}{Value of apparent flux ratio $R_\mathrm{F}$ of kpc-scale radio lobes was not shown in the previous study.  In this paper, we assume $R_\mathrm{F}\sim1.2$, which was roughly evaluated from the figure in \citet{Gliozzi:2010}.}  
\end{table*}

It should be noted that the previous discussion considers the following three caveats.  
First, the adopted threshold of $L_\mathrm{j}/\bar{n}_\mathrm{a}$ was based on tentative values of jet kinetic power and ambient density.  To clarify the FR-I/II dichotomy in kpc-scale radio structures of NLS1s, a study of dynamical interaction between the radio lobe and cocoon is crucial \citep{Kaiser:1997,Kino:2005,Ito:2008} and will be explored in future research.  

Second, an alternative possibility is that the apparent compact structures in most cases even for radio-loud NLS1s is attributed to short-lived radio sources of $<10^4$~pc ($\la10^6$~years).  VLBI images have been obtained for several compact radio-loud NLS1s, including RX~J0806.6+7248, FBQS~J1629+4007, RX~J1633.3+4718, B3~1702+457 \citep{Doi:2007,Doi:2011a,Gu:2010}, PKS~1502+036 \citep{Fey:2000,Dallacasa:1998}, and SBS~0846+513 \citep{Taylor:2005,Kovalev:2007}.  However, no clear evidence of young radio lobes has yet been detected in pc scales (cf.~\citealt{Gallo:2006} proposed the NLS1--compact steep-spectrum~(CSS) connection on the radio-loud NLS1 PKS~2004$-$447; see also \citealt{Orienti:2012}).  

Third, the estimation of black masses through the virial relationship is controversial.    
The impact of this factor on our study is discussed in the following subsection.

\subsection{Black Hole Mass Deficit?}\label{section:BHmass_deficit}
The virial mass estimation is subject to the assumption of an isotropic geometry of a broad-line region~(BLR) with random orbital inclination of clouds that are gravitationally bound to a black hole \citep{Netzer:1990}.  
%
%
The virial assumption could be violated in sources with high Eddington ratios because the outward force due to radiation pressure overcomes or compensates for gravitational attraction (\citealt{Marconi:2008}; but see also \citealt{Netzer:2009,Marconi:2009}).   
The mass deficit due to this effect indicates that BLS1s with high Eddington ratios would be preferentially misclassified into NLS1s.
However, such a situation may be inconsistent with the observed lower detection rate of extended radio emissions in NLS1s (Section~\ref{section:fractionofNLS1RG}) because large jet kinetic powers can generally be expected from such highly mass-accreting systems with larger black hole masses.  
If the geometry of BLR is flat, a small line width would be the result of a small inclination and the mass would be underestimated \citep[e.g.,][]{Decarli:2008,La-Mura:2009}.
In fact, planar BLRs in blazars have been suggested (\citealt{Decarli:2011} and references therein, but see \citealt{Punsly:2007}).  
The mass deficit due to this effect indicates that BLS1s viewed face-on would be preferentially misclassified into NLS1s.  
However, such a situation may be inconsistent with the observed lower fraction of radio-loud objects in NLS1s compared with BLS1s \citep{Komossa:2006,Zhou:2006} in the framework of the Doppler-beaming effect.  
That is, {\it these two possible mechanisms for inducing mass deficit do not manifest in the actual radio views of NLS1s as a population}.  
Small masses have been presumed for NLS1s as a class on the basis of several other methods, e.g., an accretion-disk model fit (by fitting the spectral energy distributions with the disk and corona model; \citealt{Mathur:2001}) and the analysis of X-ray variability power density spectra \citep{Nikolajuk:2004}.

From the perspective of the present study, the planar BLR may influence individual objects of NLS1s with unusually high core dominances in particular, which indicate very small viewing angles (Section~\ref{section:coredominance}).  
The black hole masses of these blazar-like NLS1s ($\ga10^7$~M$_\sun$ in a virial method) might be underestimated: their true masses may be even larger ($\ga10^8$~M$_\sun$).      
If this is the case, {\it our conclusion that NLS1s with kpc-scale radio structures have relatively large black hole masses remains}.  
We note the existence of non-beamed radio-loud NLS1s that may be inclined, which include the lobe-dominated NLS1 \object[SDSS J120014.08-004638.7]{SDSS~J120014.08$-$004638.7} with $M_\mathrm{BH} = 10^{7.4}$~M$_\sun$ (Section~\ref{section:newdetection}) and several steep-spectrum radio sources in radio-loud NLS1s \citep{Komossa:2006,Gallo:2006,Doi:2011a}; intrinsic radio and optical natures are observed in these sources.

\subsection{Kinematic Ages of Radio Sources as Clues of Growing Black Holes}\label{section:ImplicationsforBHevolution}

The ages of NLS1s with kpc-scale radio structures provide important clues of integrated activity of a rapidly growing black hole at the lower end of the AGN mass function.  We define the kinematic age as $t^\mathrm{kpc}=L^\mathrm{deproj.}/\beta^\mathrm{kpc}c$, where $L^\mathrm{deproj.}$ is the deprojected size at kpc scales assuming the same viewing angle as that at the core region ($\Phi^\mathrm{core}$).  Assuming $\beta_\mathrm{kpc}\sim0.01$--0.3 at kpc scales [Column~(7) in Table~\ref{table3}] as constant speeds of advance, jet activity should be maintained for $\ga10^7$~years to build the entire radio structures [Column~(8) in Table~\ref{table3}].  For \object[SDSS J120014.08-004638.7]{SDSS~J120014.08$-$004638.7} and \object[SDSS J145041.93+591936.9]{SDSS~J145041.93+591936.9}, strong constraints on $t^\mathrm{kpc}$ cannot be obtained because the upper limits of $\beta^\mathrm{kpc}$ are not available from their coreless and one-sided jet structures, respectively.

In comparison with Galactic black holes \citep[][for a review]{Done:2007}, the accretion disks of NLS1s with kpc-scale radio structures could be in a very high state, which is a transitional state at very high accretion rates between low/hard and high/soft states, characterized by powerful transient relativistic ejections \citep{Gliozzi:2010}.  The derived kinematic ages of $\ga10^7$~years for these NLS1s may imply the sustenance ability of the very high state on the internal secular evolution of black holes in NLS1 hosts \citep{Kormendy:2004,Orban-de-Xivry:2011}.  NLS1s are considered to be in a super-Eddington accretion state and are believed to have slim disks as optically thick advection-dominated accretion flows \citep{Abramowicz:1988,Mineshige:2000,Wang:2003}.  Black hole masses are expected to increase exponentially by factors of $\sim10$ and $\sim1000$ in case of super-Eddington mass accretion\footnote{$L_\mathrm{Edd}=\eta \dot{M}_\mathrm{Edd} c^2$, and we assume that $\eta\sim0.1$, where $L_\mathrm{Edd}$, $\eta$, and $\dot{M}_\mathrm{Edd}$ are the Eddington luminosity, mass-energy conversion efficiency, and Eddington mass accretion rate, respectively.} at $\sim10\dot{M}_\mathrm{Edd}$ for durations of $1\times10^7$~years and $3\times10^7$~years, respectively \citep{Kawaguchi:2004}.  These arguments suggest that the lifetime of an NLS1 with relativistic ejections is limited to $\sim10^7$~years.  
From the perspective of the fraction of NLS1s ($\sim10$\%) in the AGN population as well, $\sim10^7$~years in NLS1 phase is suggested from the AGN lifetime of $\sim10^8$~years \citep{Collin:2004,Kawaguchi:2004}.  
However, large-scale radio morphology would be detectable in radio sources with ages of $\ga10^7$~years after their black holes have grown to $\ga 10^7$~$\mathrm{M_\sun}$ (Section~\ref{section:escape-to-kpc}), in which the population of broad-line AGNs is dominant (Figure~\ref{figure3}).  Hence, kpc-scale radio structures may originate in a small window of opportunity during the final stage of the NLS1 phase just before altering into broad-line AGNs.  {\it Thus, the lower detection rate of kpc-scale radio structures can be attributed to the short lifetime} in addition to lower maximal jet kinetic powers due to lower-mass black holes (Section~\ref{section:escape-to-kpc}).   

The existence of large-scale radio structures in NLS1s apparently conflicts with the paradigm that radio galaxies are associated exclusively with elliptical host galaxies \citep{Veron-Cetty:2001,Hota:2011}.  Many optical and infrared investigations of host galaxies have been conducted for nearby (radio-quiet) NLS1s, revealing higher fractions of strongly barred spirals \citep{Ohta:2007} and stronger star-forming activity \citep{Sani:2010} than those of normal broad-line Seyfert galaxies.  However, the host morphologies of the detected NLS1s are not clearly distinguished on the basis of their SDSS images \citep[cf.][]{Huertas-Company:2011,Lintott:2011} because they are relatively distant.  \object[1H 0323+342]{1H~0323+342} is the only NLS1 with kpc-scale radio structures whose host morphology has been investigated because of its proximity \citep{Zhou:2007,Anton:2008,Foschini:2011}, and it exhibits a peculiar ring morphology with a circum-nuclear starburst similar to that observed in collisional ring galaxies \citep{Anton:2008}.  The study of hosts for radio-loud NLS1s is crucial for testing the paradigm of radio galaxies in ellipticals, even at rapidly growing phases of supermassive black holes.

\section{Summary}\label{section:summary}
The detection of extended radio emissions in radio-loud and $\gamma$-ray emitting NLS1s is crucial for understanding the jet activity of the NLS1 class in the framework of the unified scheme of radio-loud AGNs.  This study is summarized in the following points:

\begin{itemize}
\setlength{\itemsep}{0pt}

\item We found extended radio structures in three NLS1s from VLA FIRST images (Sections~\ref{section:data} and \ref{section:newdetection}), which increased the number of known radio-loud NLS1s with kpc-scale radio structures to six.  

\item The detection rate of extended radio emission in NLS1s is lower than that in broad-line AGNs with a statistical significance of 5\% (Section~\ref{section:fractionofNLS1RG}).

\item Four of the six NLS1s (\object[PMN J0948+0022]{PMN~J0948+0022}, \object[FBQS J1644+2619]{FBQS~J1644+2619}, \object[1H 0323+342]{1H~0323+342}, and \object[PKS 0558-504]{PKS~0558$-$504}) exhibit two-sided radio structures in kpc scales, one-sided jet morphology in pc scales (Sections~\ref{section:advancingspeed}), and unusually high core dominances, as observed in most blazers.  These characteristics are suggested by the concept of a Doppler-boosted core and sufficiently decelerated radio lobes (Section~\ref{section:coredominance}).

\item We also detected a two-sided lobe-dominated structure, indicating a radio galaxy, in an NLS1 (\object[SDSS J120014.08-004638.7]{SDSS~J120014.08$-$004638.7}; Section~\ref{section:newdetection}).    

\item Thus, the NLS1s with kpc-scale radio structures can be understood in the framework of the unified scheme of radio-loud AGNs that considers radio galaxies as non-beamed parent populations of blazars (Section~\ref{section:coredominance}).  

\item Five of the six NLS1s are driven by relatively high-mass black holes of $\ga 10^{7}\ \mathrm{M_\sun}$ in contrast to $\la 10^{7}\ \mathrm{M_\sun}$ of the NLS1 population (Section~\ref{section:escape-to-kpc}).

\item The five NLS1s with $\ga 10^{7}\ \mathrm{M_\sun}$ have extended radio luminosities equivalent to jet kinetic powers of $L_\mathrm{j} \ga 10^{44}$~erg~s$^{-1}$, which are sufficient to escape from host galaxies (Section~\ref{section:escape-to-kpc}).  
 
\item Two-sidedness in kpc scales requires expansion rates of $\sim0.01c$--$0.3c$, resulting in kinematic ages of $\ga10^7$~years for deprojected radio sizes (Section~\ref{section:ImplicationsforBHevolution}).  

\item The lower detection rate of extended radio emissions in the NLS1 population may be attributed to (1)~lower-mass black holes of $\la10^{7}$~$\mathrm{M_\sun}$, which generate lower maximal jet kinetic powers (Section~\ref{section:escape-to-kpc}), and (2)~a limited lifetime of $\sim10^7$~years in NLS1 phase, which is too short for the expansion of radio structures (Section~\ref{section:ImplicationsforBHevolution}), in contrast with $>10^{7}$~$\mathrm{M_\sun}$ and $\sim10^8$~years for broad-line AGNs. 

\end{itemize}

\bigskip 

\acknowledgments
We are grateful to the anonymous referee for offering constructive comments that have contributed in substantially improving this paper.  In addition, we thank Luigi Foschini for his valuable comments.  In the present study, we used NASA's Astrophysics Data System Abstract Service and the NASA/IPAC Extragalactic Database~(NED), which is operated by the Jet Propulsion Laboratory.  The VLA is operated by the National Radio Astronomy Observatory, which is a facility of the National Science Foundation operated under cooperative agreement by Associated Universities, Inc.  This study is supported in part by a Grant-in-Aid for Scientific Research~(C; 21540250, AD, and B; 24340042, AD), Japan Society for the Promotion of Science~(JSPS), the Ministry of Education, Culture, Sports, Science and Technology (MEXT) Research Activity Start-up (2284007, NK), and the Center for the Promotion of Integrated Sciences~(CPIS) of Sokendai.





\end{document}